# Anomalous attenuation of extraordinary waves in ionosphere heating experiments: experimental results of 2000-2001

N.A. Zabotin, G.A. Zhbankov and E.S. Kovalenko

*Rostov State University, Rostov-on-Don, Russia*

V.L. Frolov, G.P. Komrakov, N.A. Mityakov, E.N. Sergeev

*NIRFI, Nizhni Novgorod, Russia*

## Abstract

Multiple scattering from artificial random irregularities HF-induced in the ionosphere F region causes significant attenuation of both ordinary and extraordinary radio waves together with the conventional anomalous absorption of ordinary waves due to their conversion into the plasma waves. To study in detail features of this effect, purposeful measurements of the attenuation of weak probing waves of the extraordinary polarization have been performed at the Sura heating facility. Characteristic scale lengths of the involved irregularities are ~0.1-1 km across the geomagnetic field lines. To determine the spectral characteristics of these irregularities from the extraordinary probing wave attenuation measurements, a simple procedure of the inverse problem solving has been implemented and some conclusions about the artificial irregularity features have been drawn. Theory and details of experiments have been stated earlier. This paper reports results of two experimental campaigns carried out in August 2000 and June 2001 under support of Russian Foundation for Basic Research (grants No. 99-02-17525 and No. 01-02-31008). Particularity of these experiments consisted in using of lower heating power (20-80 MW ERP). Regular character of the multiple scattering effects has been confirmed.

## Results and discussion

Brief statement of theory with necessary references, detailed description of our experimental technique, and results of experimental campaign of year 1999 are contained in our earlier electronic publication [1]. Journal version of that paper is currently under review at the Radio Science. In the present paper we shall use the same methods of presentation of our results, without repetition of detailed explanations.

## a) August 2000 campaign

The experiments at the Sura heating facility were carried out on August 22 and 23, 2000 in evening and night hours when the linear absorption of radio waves in the ionosphere D and E regions due to electron-ion collisions was negligible. Ionosphere heating was provided by coherent HF radiation of ordinary polarization of single transmitter with output power of 250 kW. With account of the antenna gain the effective radiated power (ERP) was 20 MW. Such relatively low power level was chosen intentionally, in contrast with our experiments of year 1999, to avoid effect of large-scale irregularity amplification. Duration of each heating cycle was 10 minutes including the active phase of 3 minutes duration (heater on) and off-time of 7 minutes. Overall duration of the measurements included into processing was about of 4.3 hours (26 heating cycles) and 3.2 hours (19 heating cycles) on 22 and 23 August, respectively. Amplitudes of X-mode probing waves were recorded at 7 frequencies. Figs. 1 and 2 show amplitudes of the heating wave and the probe waves averaged over all available heating cycles. They give clear evidence of presence of the anomalous attenuation (AA) effect in waves of both polarization.

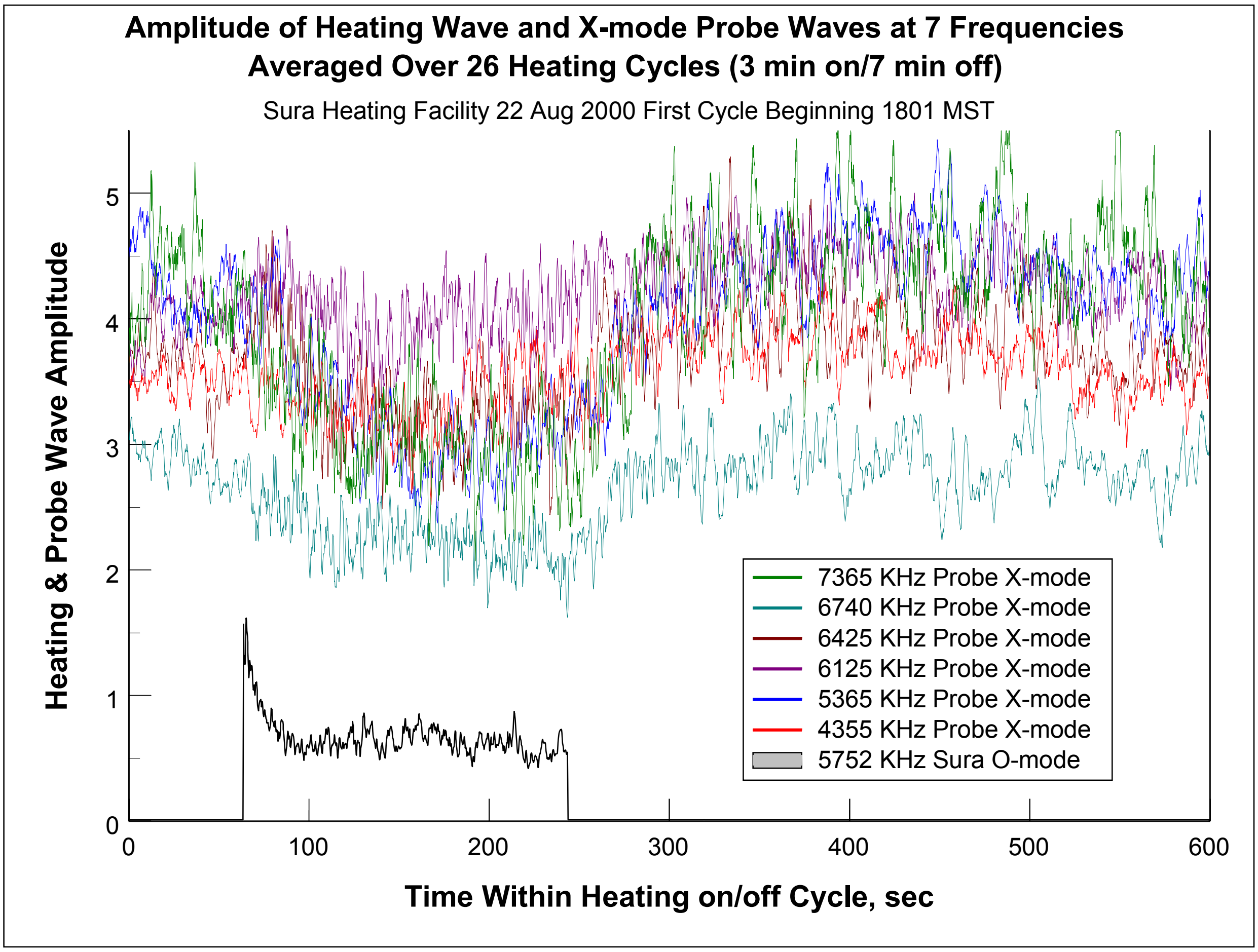

Figure 1.

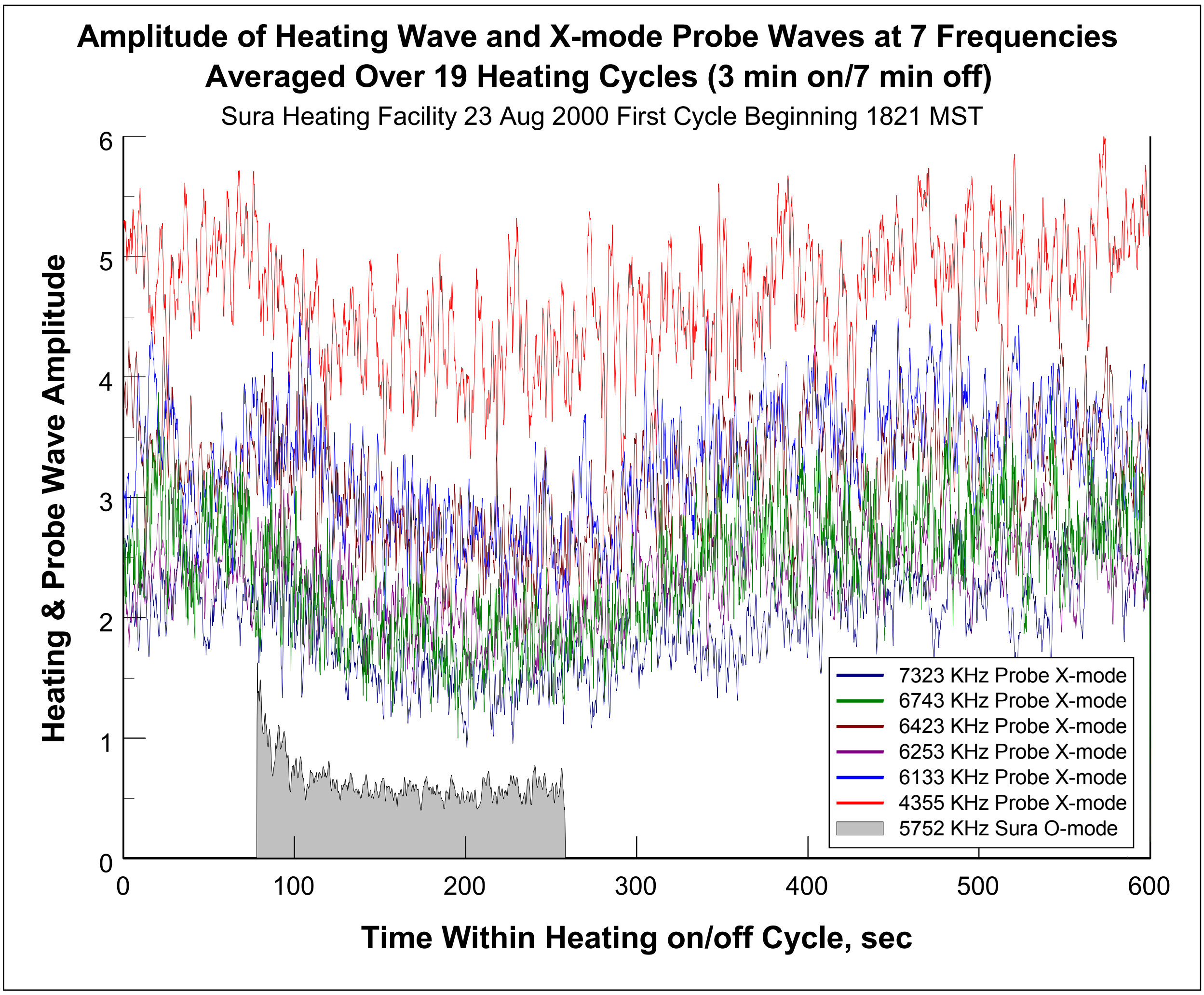

Figure 2.

This effect is characterized by the relation of the signal intensity before the heating beginning and at the end of the heating on period, expressed in dB. Averaged frequency dependence of this quantity for the X-waves is shown in Fig. 3.

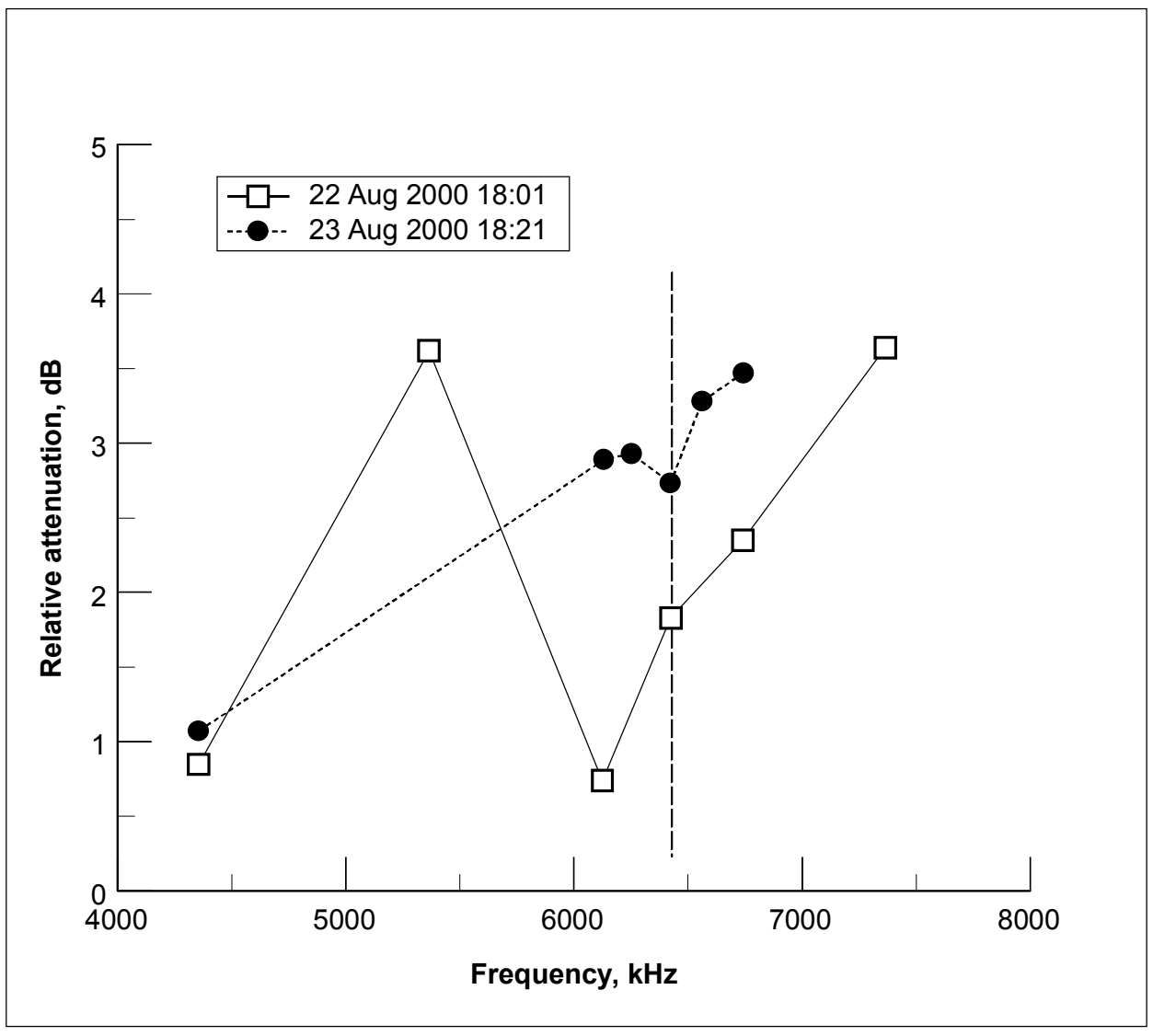

Figure 3.

One characteristic feature of this dependence is weakening of the AA effect right at (or slightly below) the probe wave frequency that corresponds to reflection near the center altitude of the heating region (it is marked by vertical dashed line). Characteristic times of the effect development and relaxation (after the heater off) are shown in Figs. 4 and 5. The time of development has tendency to increase when the probe waves are reflected closer to the heating region. The time of relaxation is approximately constant for all frequencies.

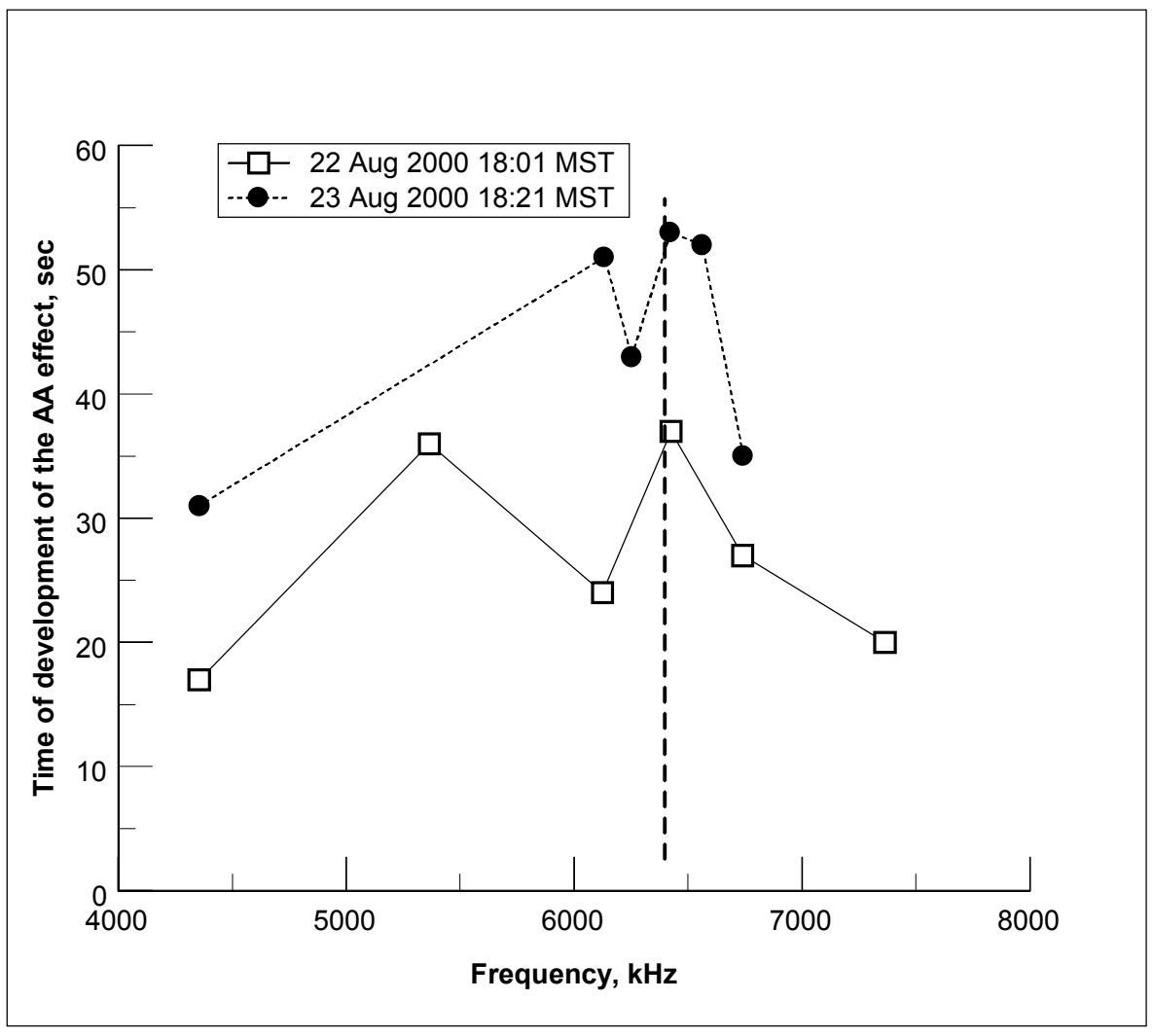

Figure 4.

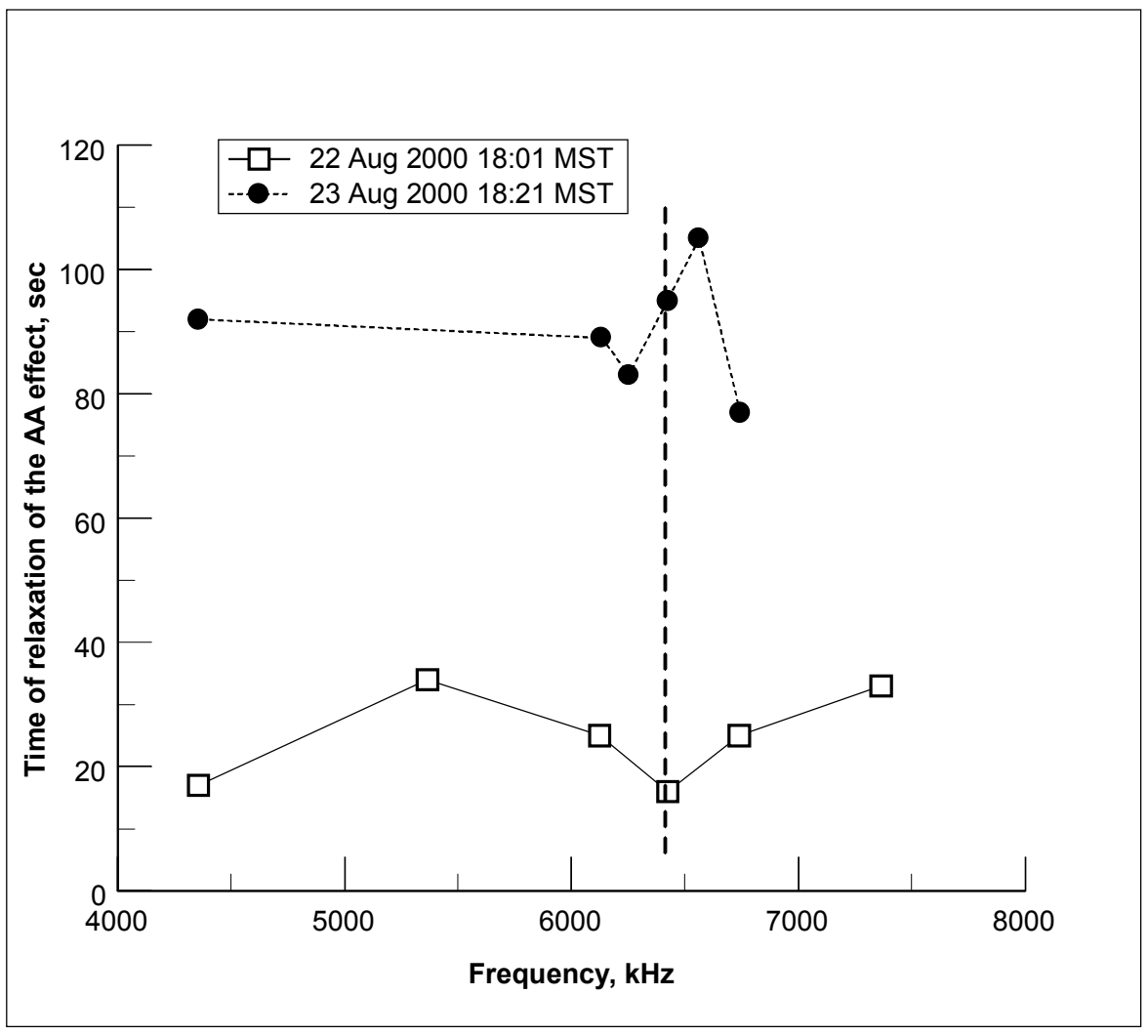

Figure 5.

The last two figures for this series, 6 and 7, present results of solving the inverse problem for the irregularity amplitude $\Delta N/N$ for several supposed values of the background irregularity level (shown in percent near curves). Note relative decrease of the irregularity amplitude near the heating region.

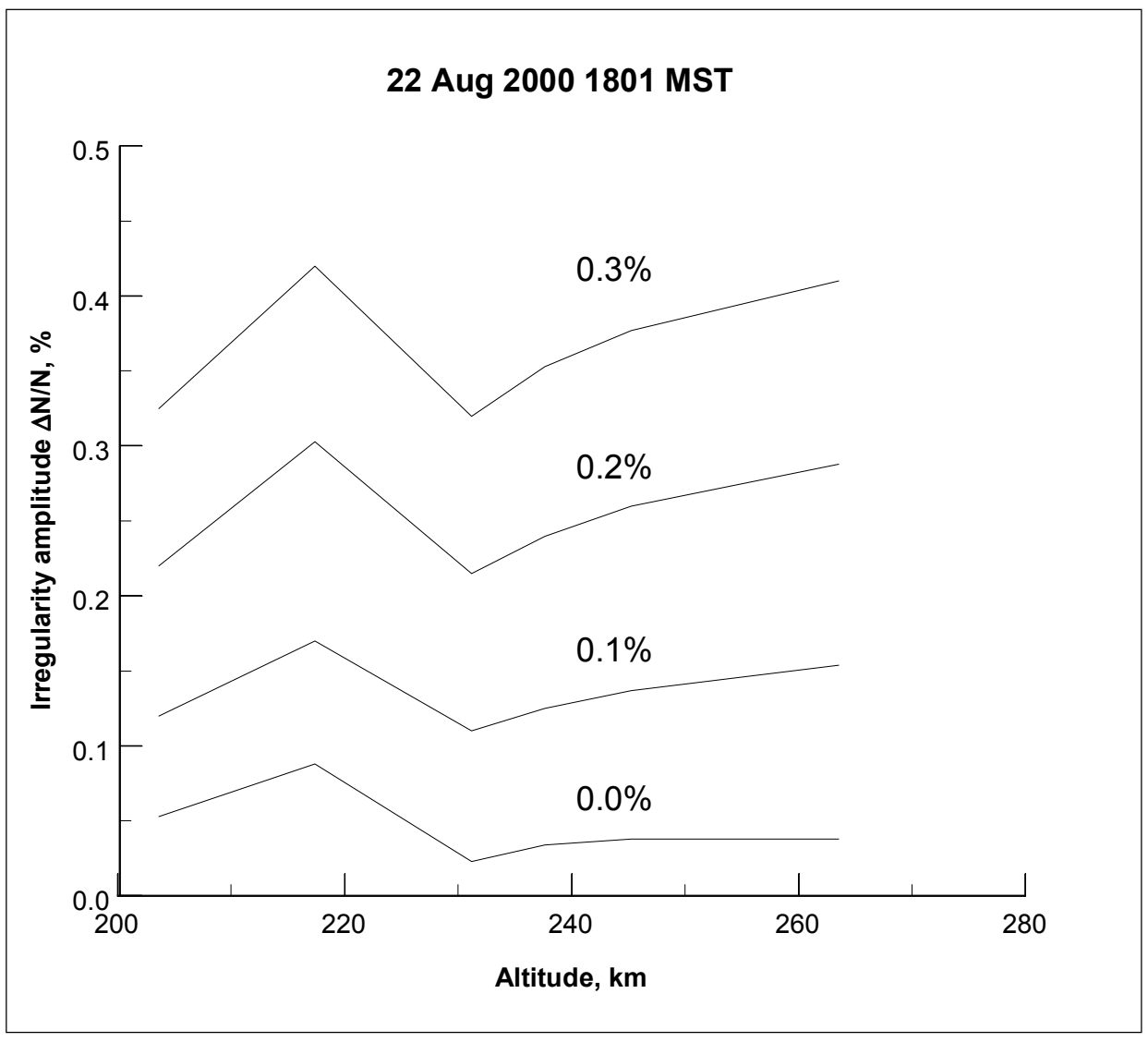

Figure 6.

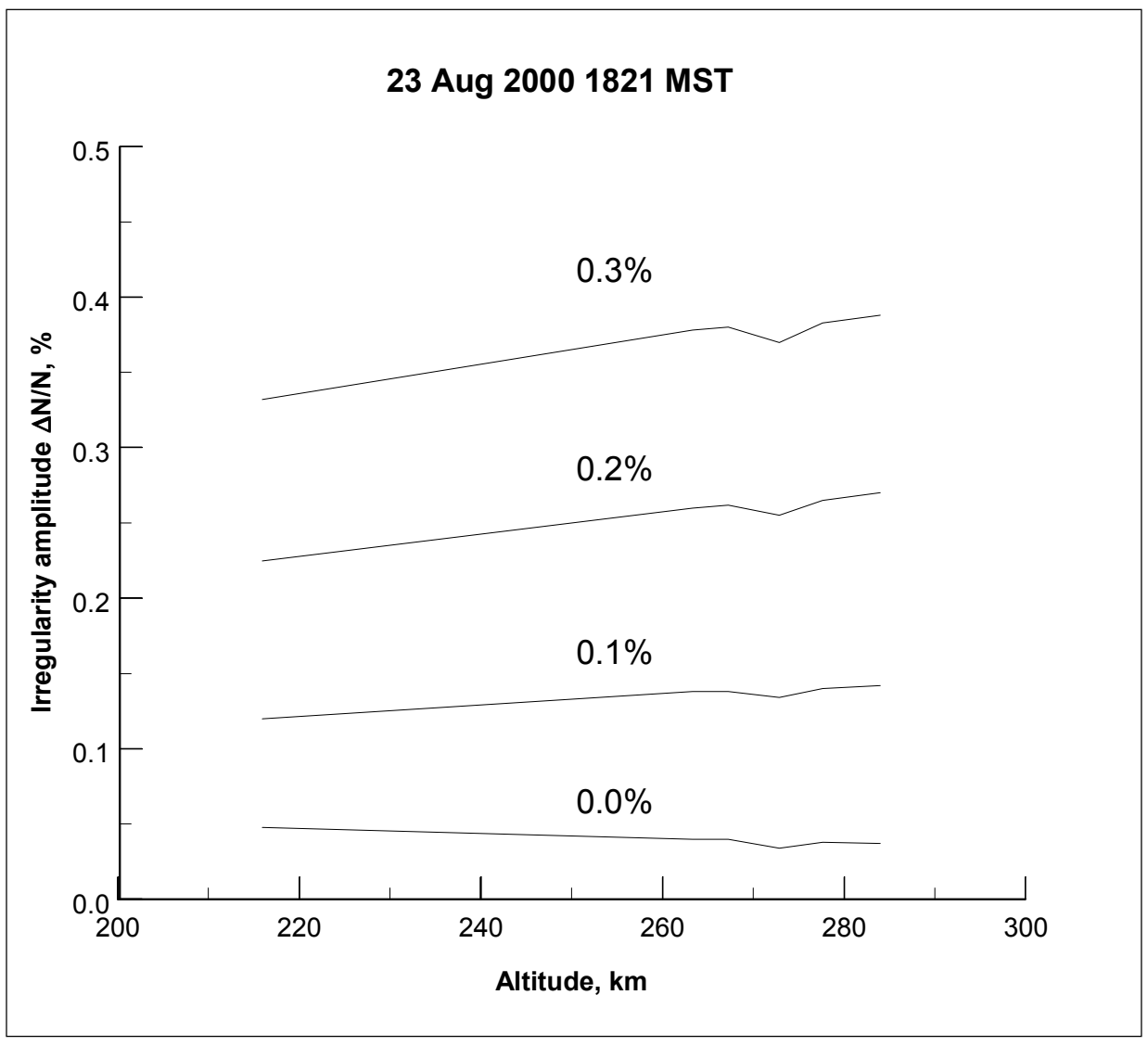

Figure 7.

## b) June 2001 campaign

The experiment at the Sura heating facility was carried out on June 1, 2001, also in evening and night hours. The whole period was divided for three intervals beginning at 17:22, 18:23, and 19:32 MST. During these intervals different heating power values and different frequency sets were used. The effective radiated power was 20 MW for the first and the third intervals and 80 MW for the second one. Duration of each heating cycle was 10 minutes including the active phase of 2 minutes and off-time of 8 minutes. Amplitudes of X-mode probing waves were recorded at 7 frequencies. Figs. 8, 9 and 10 show amplitudes of the heating ordinary wave and the probe waves averaged over all available heating cycles. The anomalous attenuation effect is also evident in this case.

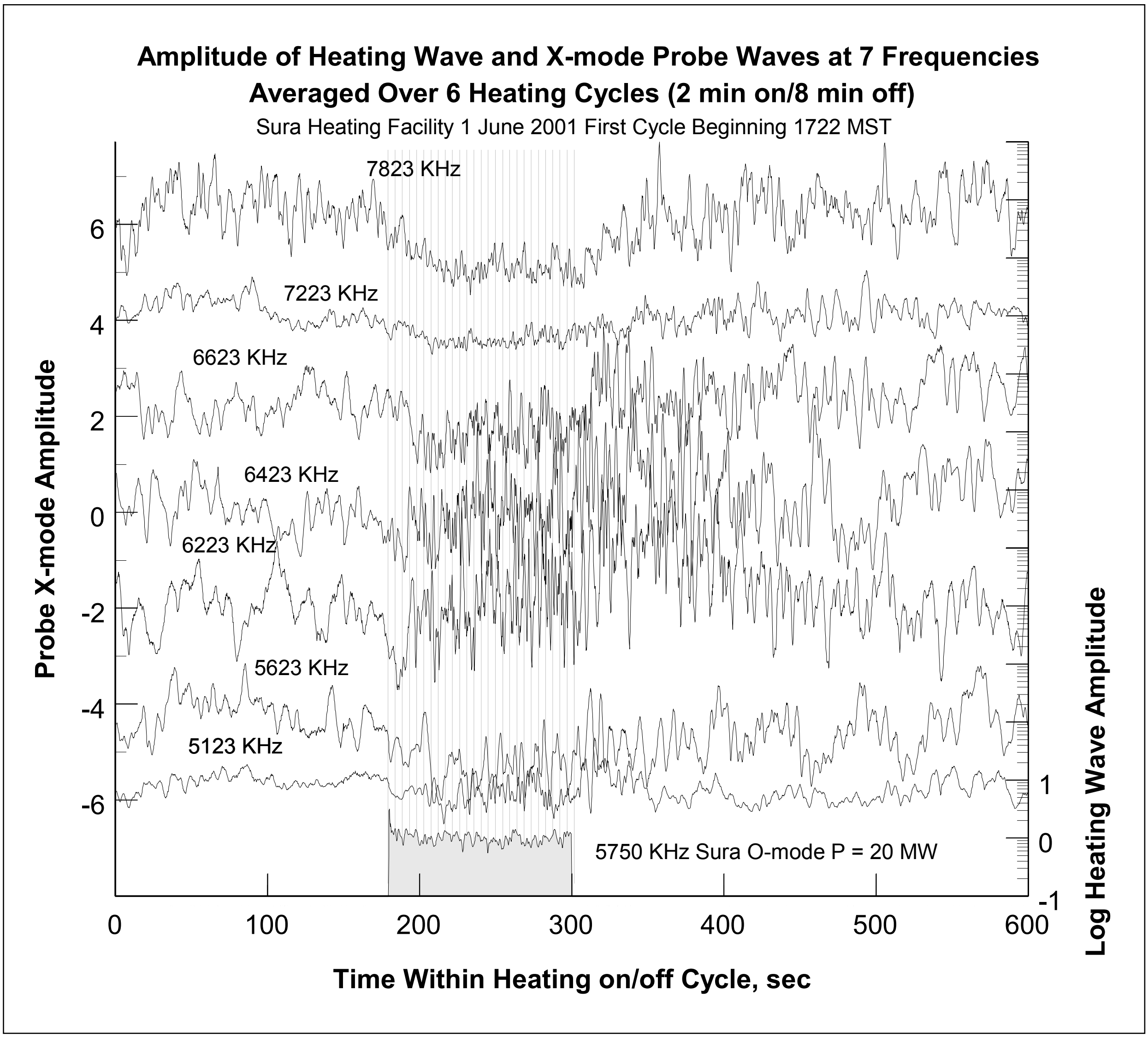

Figure 8.

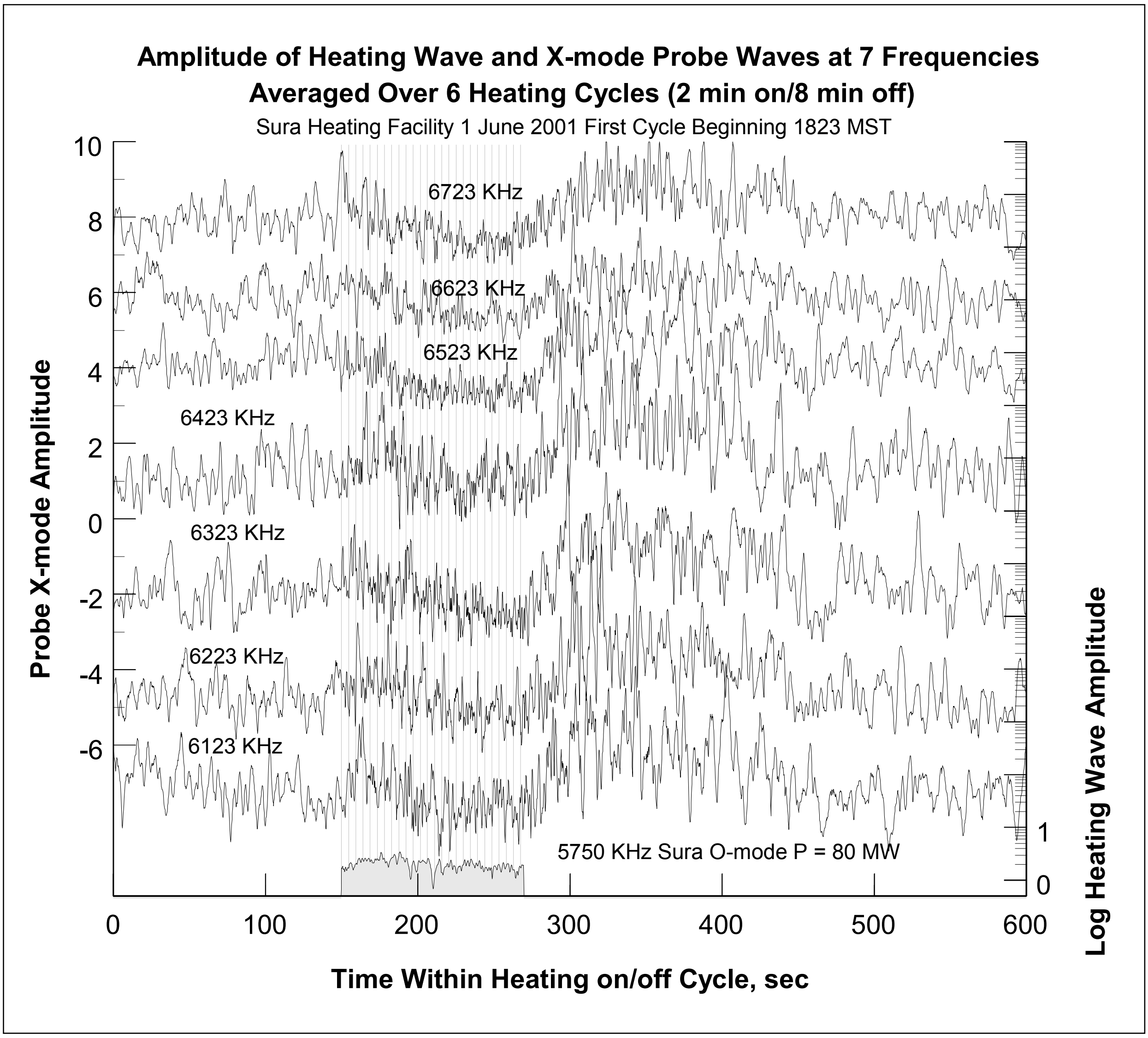
Amplitude of Heating Wave and X-mode Probe Waves at 7 Frequencies
Averaged Over 6 Heating Cycles (2 min on/8 min off)
Sura Heating Facility 1 June 2001 First Cycle Beginning 1823 MST
6723 KHz
6623 KHz
6523 KHz
6423 KHz
6323 KHz
6223 KHz
6123 KHz
5750 KHz Sura O-mode P = 80 MW
Probe X-mode Amplitude
Log Heating Wave Amplitude
10
8
6
4
2
0
-2
-4
-6
1
0
0
100
200
300
400
500
600
Time Within Heating on/off Cycle, sec

Figure 9.

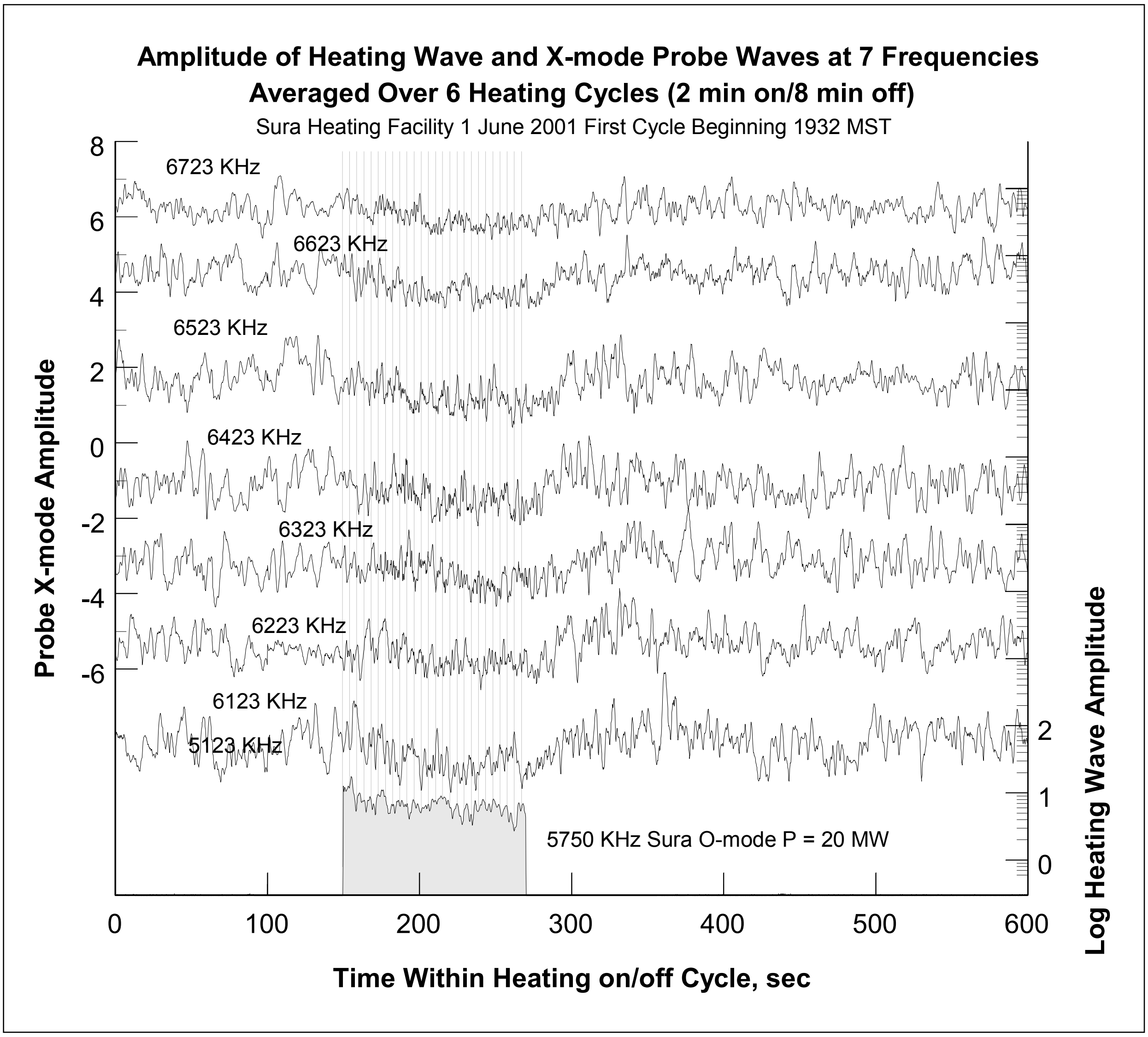
Amplitude of Heating Wave and X-mode Probe Waves at 7 Frequencies
Averaged Over 6 Heating Cycles (2 min on/8 min off)
Sura Heating Facility 1 June 2001 First Cycle Beginning 1932 MST
6723 KHz
6623 KHz
6523 KHz
6423 KHz
6323 KHz
6223 KHz
6123 KHz
5123 KHz
5750 KHz Sura O-mode P = 20 MW
Probe X-mode Amplitude
8
6
4
2
0
-2
-4
-6
Log Heating Wave Amplitude
2
1
0
0
100
200
300
400
500
600
Time Within Heating on/off Cycle, sec

Figure 10.

Averaged frequency dependence of the AA value for the X-waves is shown in Fig. 11. It is seen again that effect is especially weak within the heating region (corresponding frequency is marked by vertical dashed line).

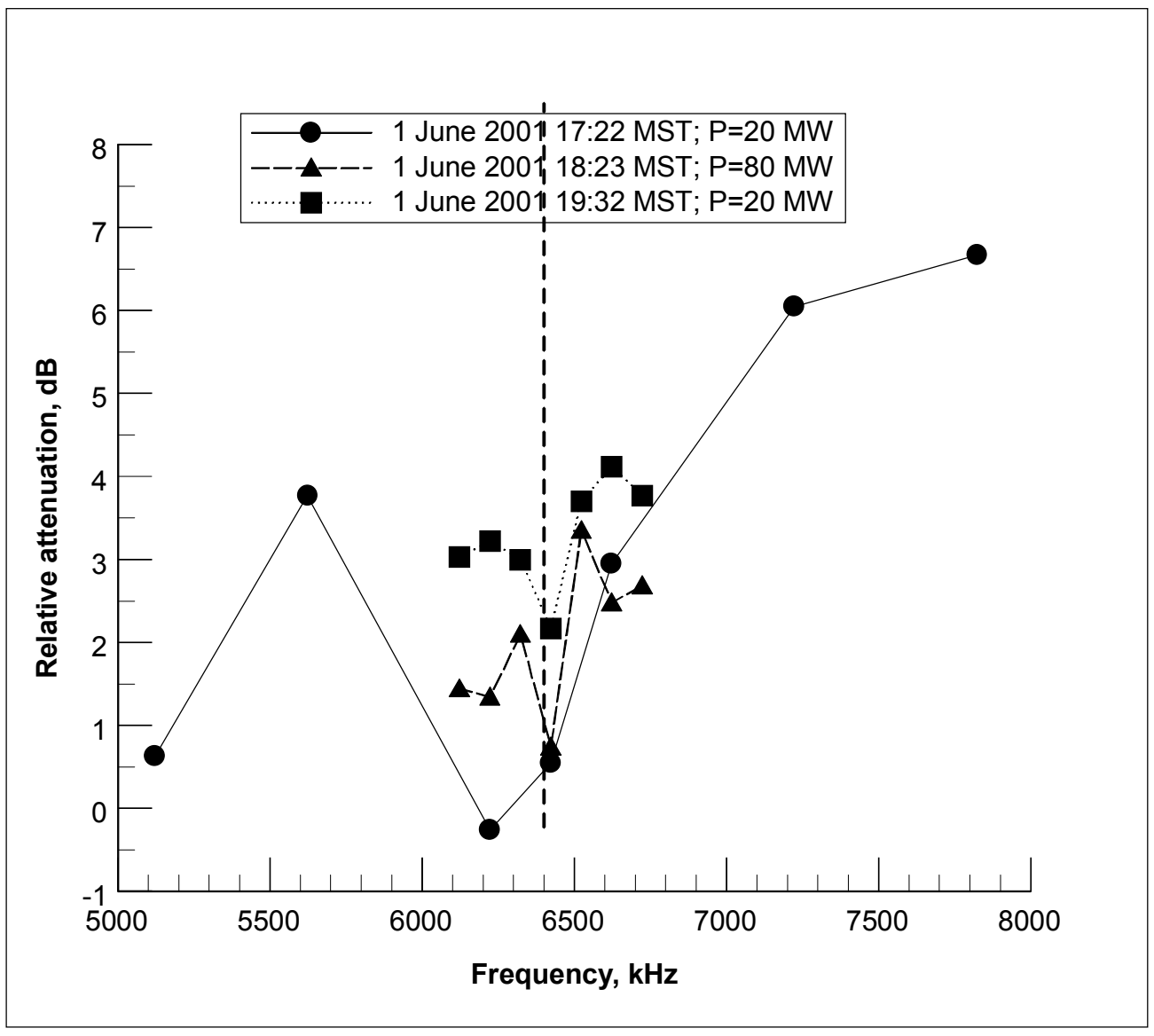

Figure 11.

Figure 12 presents results on the time of development of the AA effect.

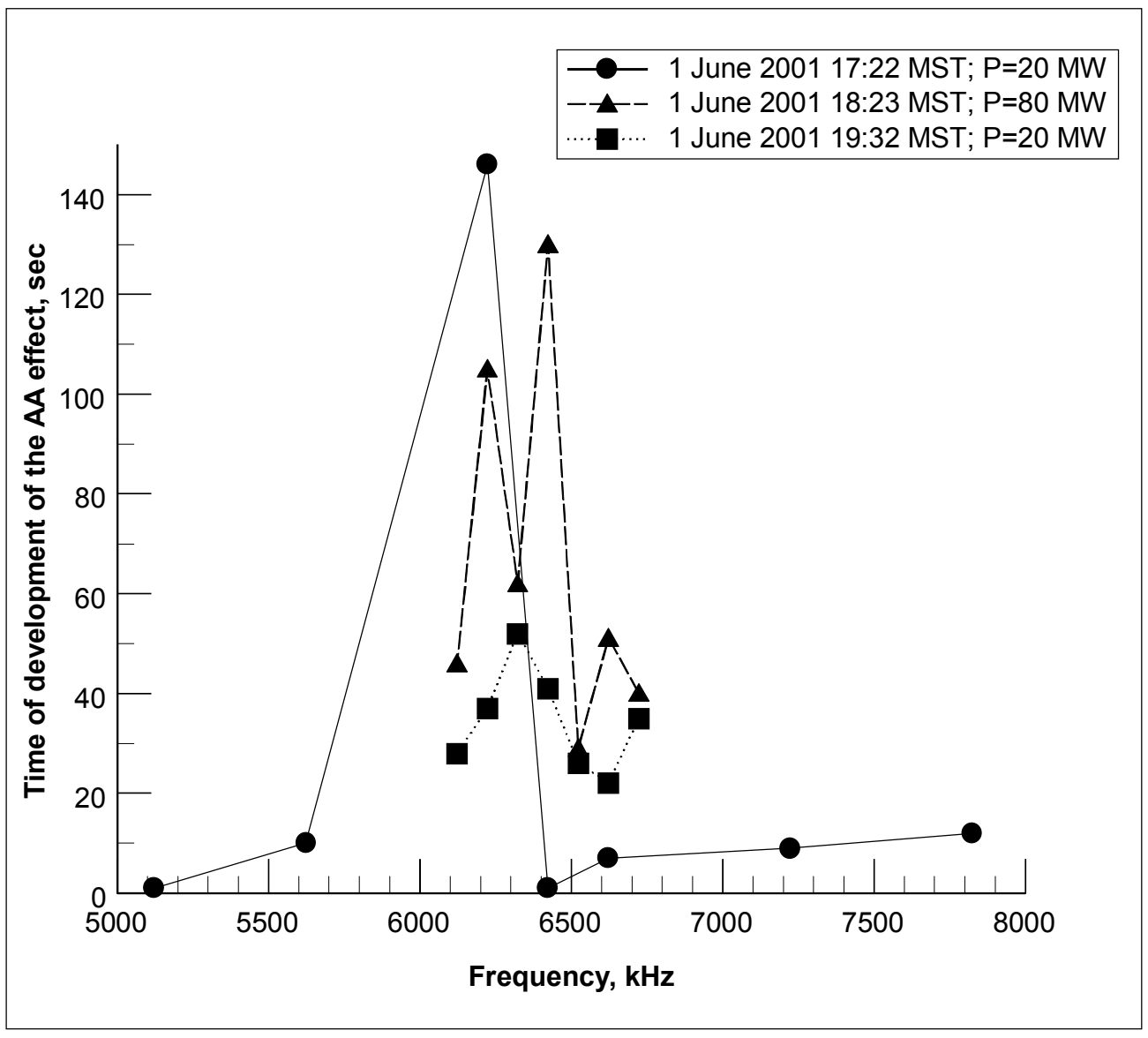

Figure 12.

The following set of illustrations (Figs. 13-18) presents results of the inverse problem solving for the irregularity amplitude $\Delta N$ and its dimensionless value $\Delta N/N$, for several supposed values of the background irregularity amplitude and broad altitude range. Common feature of all dependencies is relative drop of the irregularity amplitude in the heating region vicinity (corresponding altitude is marked by vertical dashed line).

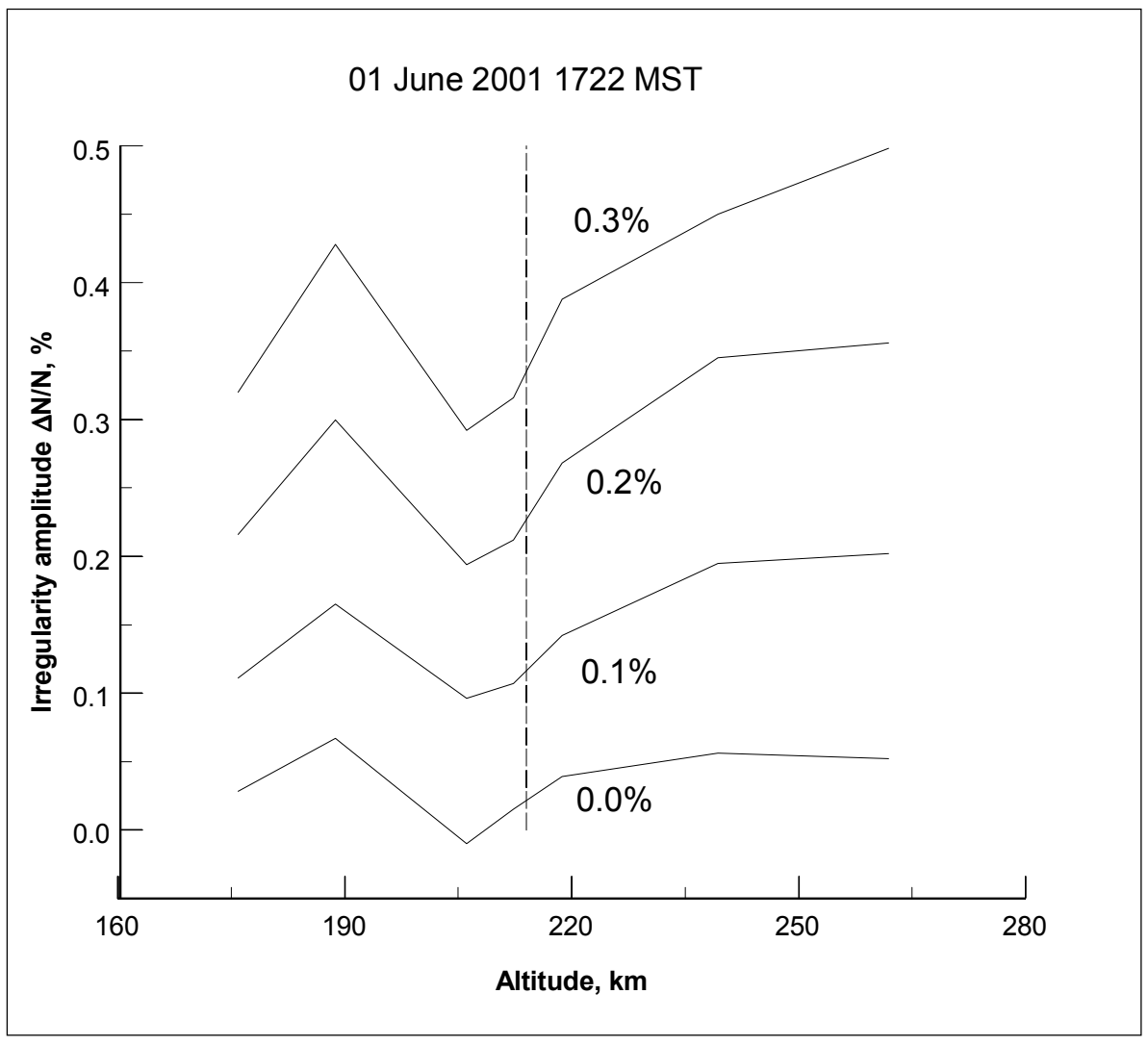

Figure 13.

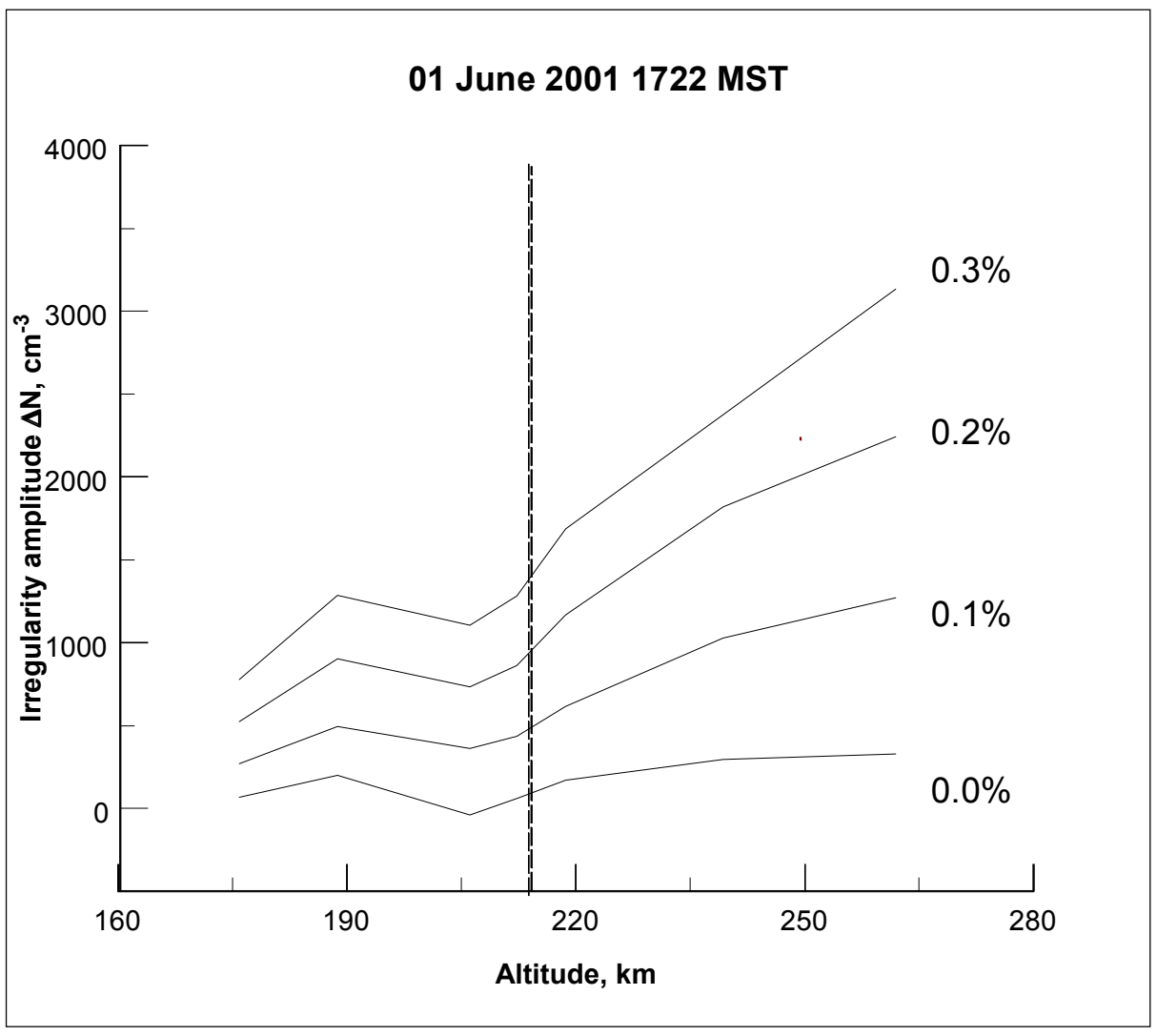

Figure 14.

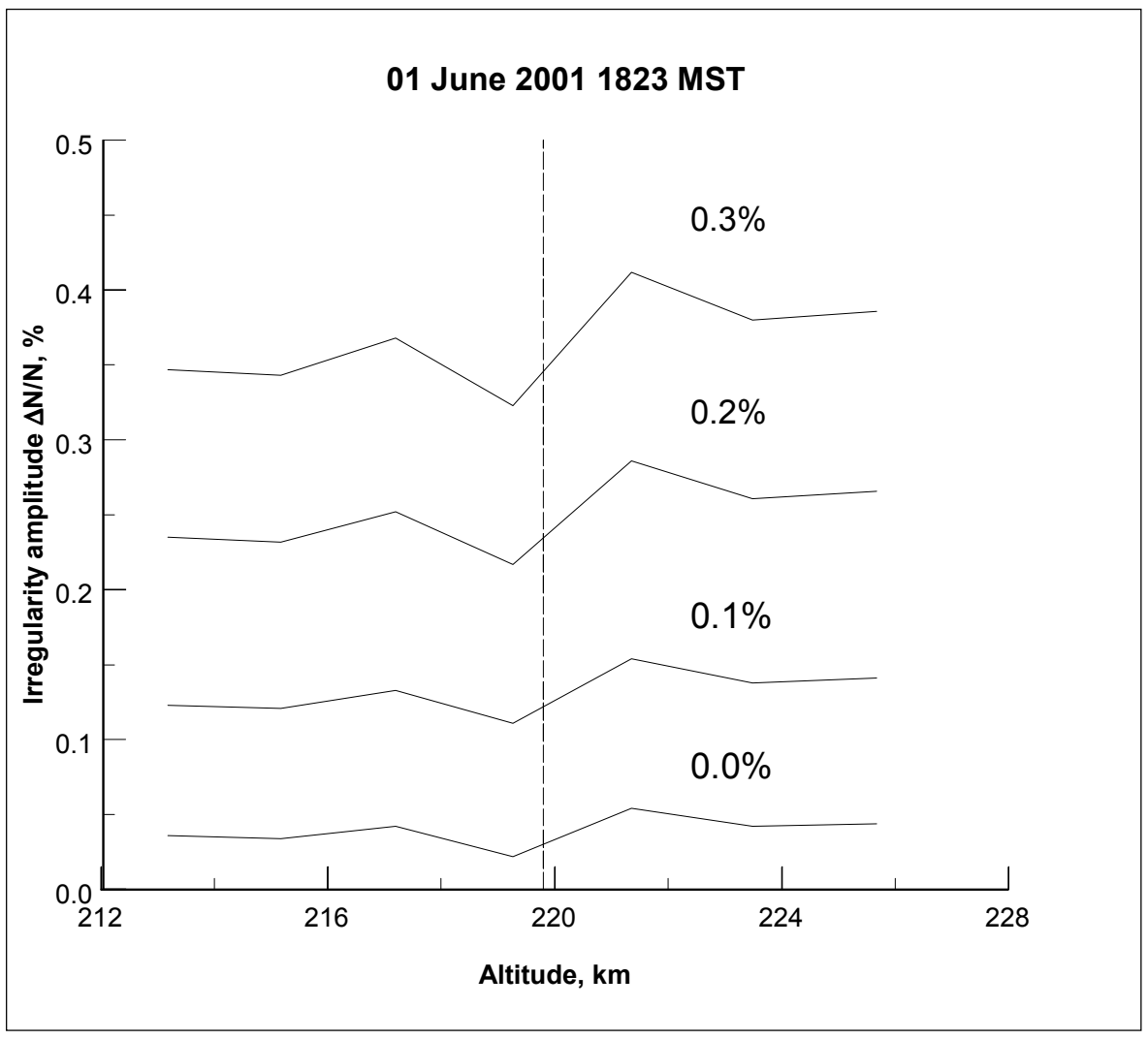

Figure 15.

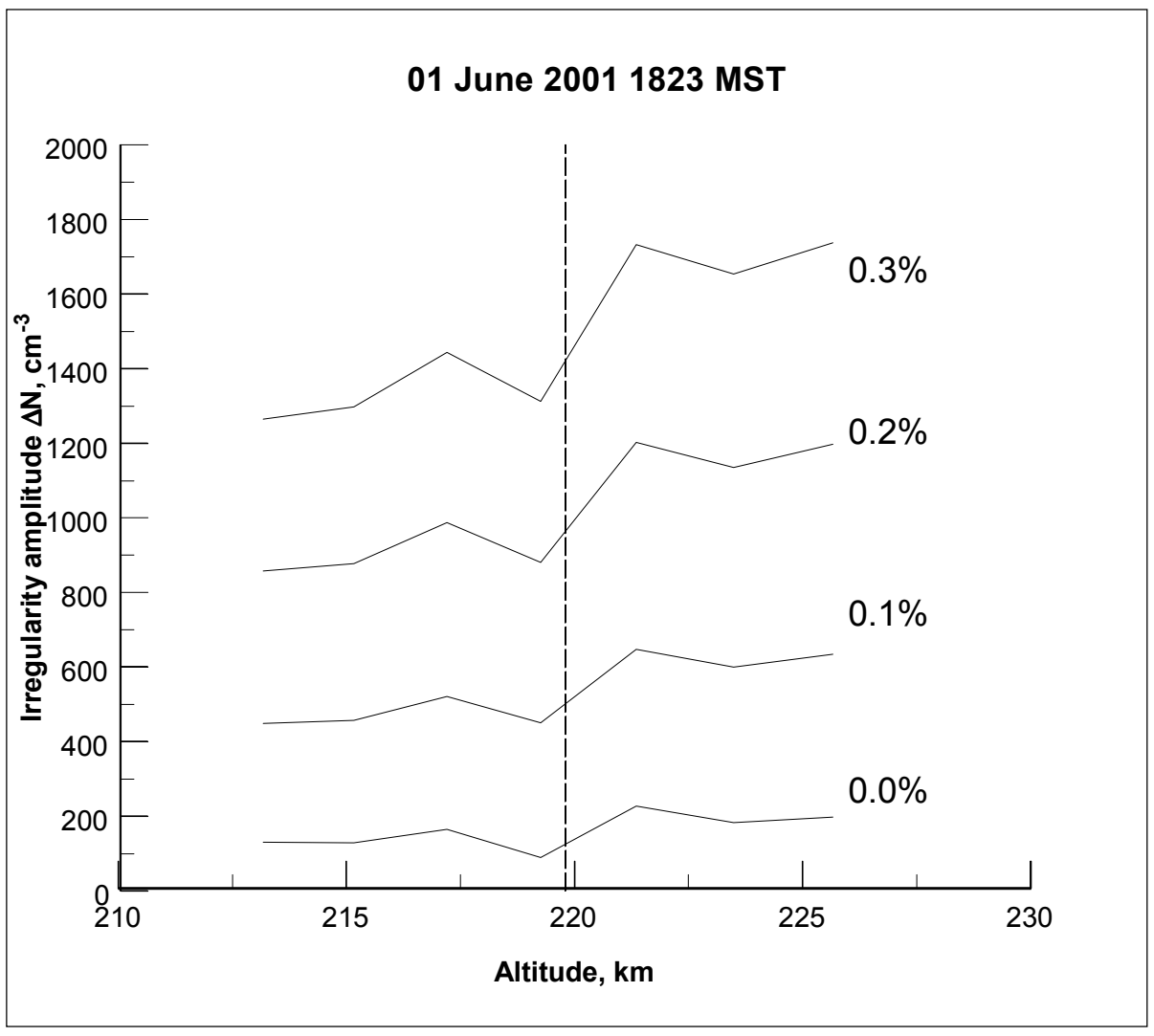

Figure 16.

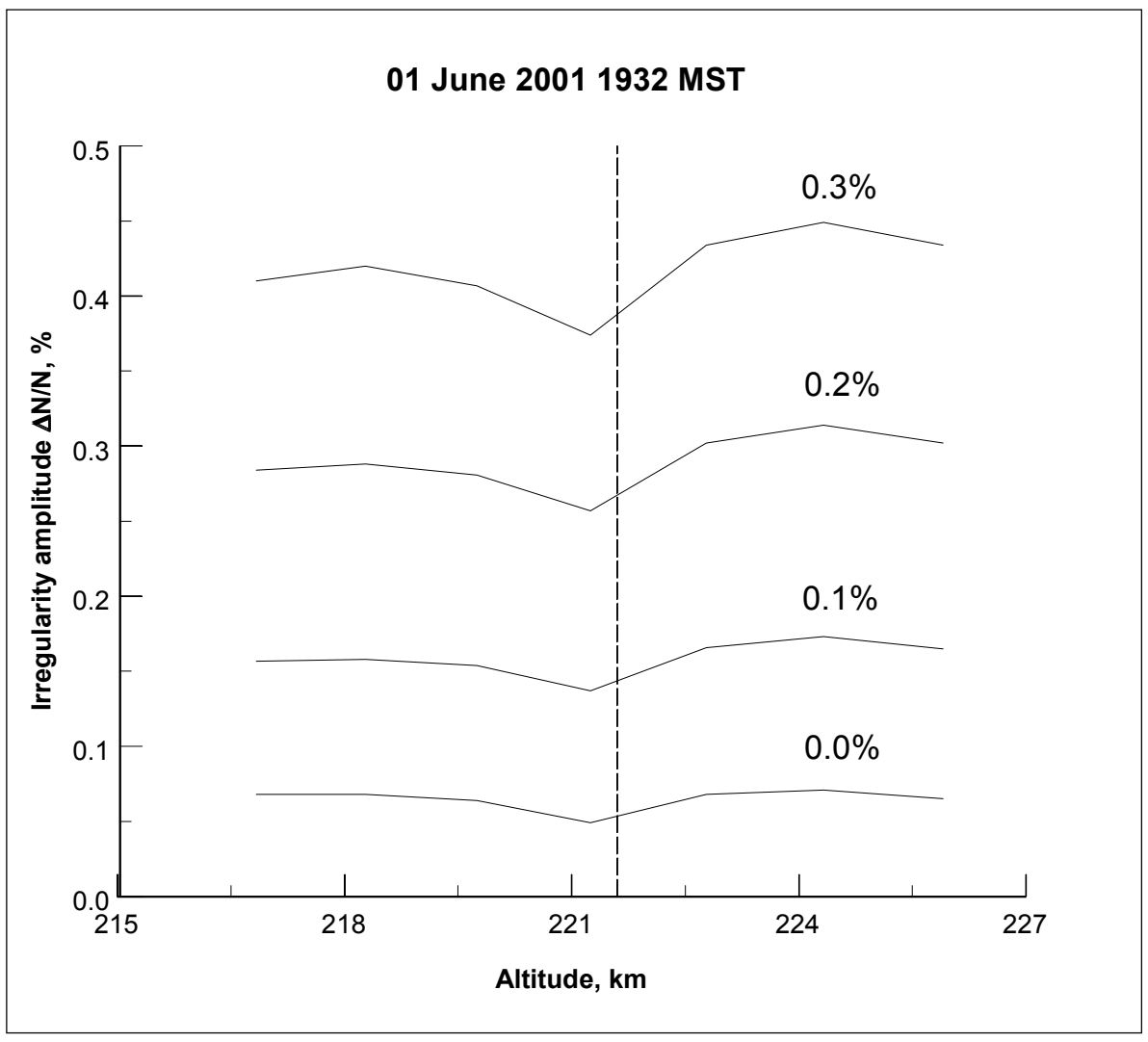

Figure 17.

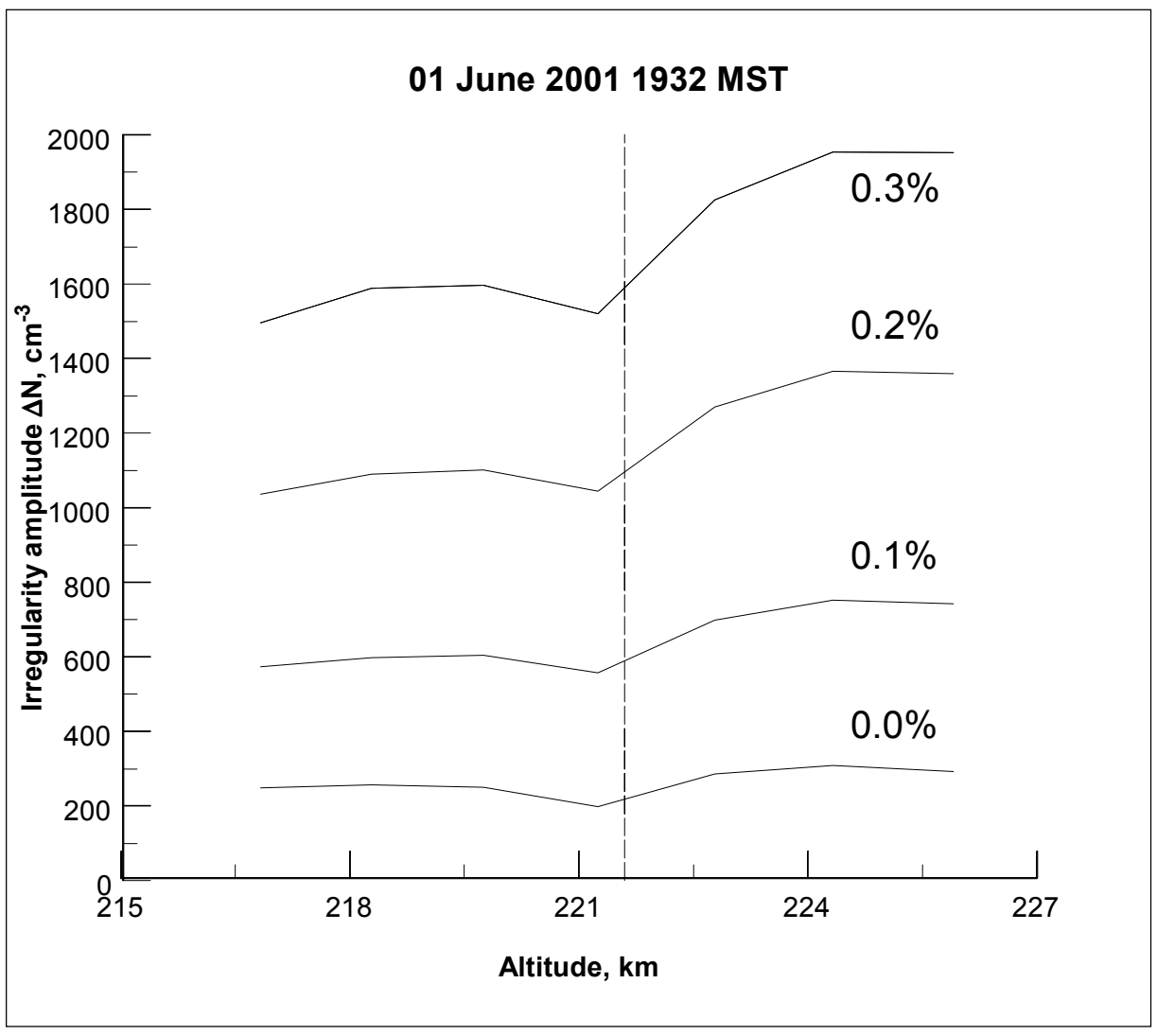

Figure 18.

## Conclusion

1) Results of the three experimental campaigns at the Sura heating facility confirm constant presence of the AA effect for the probe waves of extraordinary polarization. The only valid mechanism of this effect is multiple scattering.
2) The AA effect is observed for broad frequency range what means that heating causes irregularity growth in a broad altitude range.
3) One paradoxical phenomenon is observed: the irregularity growth is more effective outside of the heating region. We leave discussion of possible physical mechanisms of this phenomenon for another publication.

***Acknowledgements.*** This work has been supported by the Russian Foundation for Basic Research grants No. 99-02-17525 and No. 01-02-31008.